\documentclass[10pt,letterpaper]{article}
\usepackage{cogsci}
\cogscifinalcopy
\usepackage{graphicx}
\usepackage{verbatim}
\usepackage{pslatex}
\usepackage{apacite}
\usepackage{soul}
\usepackage[normalem]{ulem} 
\usepackage[T1]{fontenc}
\usepackage[sort&compress]{natbib}
\usepackage{amsmath}

\begin{document}

\title{How the Stroop Effect Arises from Optimal Response Times\\in Laterally Connected Self-Organizing Maps}

\author{
{\large \bf Divya Prabhakaran (divyapr@cs.utexas.edu)\footnotemark[1]},
{\large \bf Uli Grasemann (uli.grasemann@gmail.com)\footnotemark[1]},\\[0.25ex]
{\large \bf Swathi Kiran (kirans@bu.edu)\footnotemark[2]}, {\large \bf Risto Miikkulainen (risto@cs.utexas.edu)\footnotemark[1]}\\[0.35ex]
$^1$The University of Texas at Austin, Austin, TX 78712 USA\\[0.25ex]
$^2$Boston University, Boston, MA 02215 USA
}
\maketitle

\begin{abstract}
The Stroop effect refers to cognitive interference in a color-naming task: When the color and the word do not match, the response is slower and more likely to be incorrect. The Stroop task is used to assess cognitive flexibility, selective attention, and executive function. This paper implements the Stroop task with self-organizing maps (SOMs): Target color and the competing word are inputs for the semantic and lexical maps, associative connections bring color information to the lexical map, and lateral connections combine their effects over time. The model achieved an overall accuracy of 84.2\%, with significantly fewer errors and faster responses in congruent compared to no-input and incongruent conditions. The model's effect is a side effect of optimizing speed-accuracy tradeoffs, and can thus be seen as a cost associated with overall efficient performance. The model can further serve studying neurologically-inspired cognitive control and related phenomena. 

\end{abstract}

\section{Introduction}

Traditionally used in neuropsychological assessments, the Stroop Color and Word Test (SCWT) involves subjects reading words written in colored ink and naming colors (color-word pairs). The SCWT uses congruent and incongruent conditions of color-word naming, where incongruent conditions are represented by mismatched color-words, such as the word "red" in green ink \citep{stroop1935studies}. The Stroop effect refers to cognitive interference in the disruption of processing speed and naming when a stimulus (semantic word) interferes with the concurrent processing of a second stimulus (color), resulting in slower response times and an increased number of naming errors with incongruent stimuli \citep{10.3389/fpsyg.2017.00557}. Stroop task conditions are widely used to assess cognitive control as a measure of processing speed, selective attention, cognitive flexibility, and executive function \citep{JENSEN196636}. 

Behavioral studies of the Stroop effect, focused on summary measures such as reaction times and error rates, are not well suited for identifying the underlying mechanisms of interference. Similarly, studies of brain region activations alone cannot reveal precise information about how Stroop interference occurs: control and interference can occur at multiple time points from task initiation to later stages of response production. Computational models can lead to such insights. For instance, existing models \citep{Cohenparallelprocessing, Stroopindividualdiffs, DNN_CBF_Stroop,KAPLAN20071414, stafford2005basal, verguts2017binding, herd2006neural} produce the Stroop effect based on top-down control mechanisms where processing pathways are modulated based on task demands. However, recent critique suggests that attention driven by the salience and relevance of stimuli could cause Stroop interference based on bottom-up, input-driven processes \citep{algom2019reclaiming}.

The study described in this paper addresses this gap between top-down and bottom-up mechanisms. It demonstrates how the Stroop effect arises mechanistically and naturally from the model's underlying structure of cortically-inspired, self-organized lexical and perceptual representations. Importantly, the model explains not just how the Stroop interference occurs, but also why it represents an optimization of efficient language and perceptual processing. It can thus lead to general insights into executive function and cognitive control by highlighting how attentional constraints and stimulus-driven processing supplement or constrain these higher-order cognitive processes.

\section{Background}

The neural circuitry underlying the Stroop effect is relatively well understood, leading to several efforts to model it computationally.

\subsection{Cognitive Neuroscience Models}
The Stroop effect is believed to arise from processing in the cortico-striatal network. This network is involved in efficient language processing through the regulation and allocation of cognitive resources such as verbal attention and working memory \citep{JACQUEMOT2021104785}. The cortex-basal ganglia-thalamus loop is a crucial part of information routing and selection for cognitive control. The basal ganglia (BG) acts as a cognitive selection mechanism by optimizing information gating, as it receives input from the striatum containing processed signals from the cortex, and projects output through the thalamus back to the cortex \citep{stafford2005basal}. The left caudate nucleus of the BG plays an important role in suppressing competing alternatives in the indirect pathway and facilitating selected cortical representations, as well as language selection and control \citep{abutalebi2007bilingual, friederici2003role}.

Neuroimaging studies that focused on goal-directed behavior in the prefrontal cortex (PFC) component of the cortico-striatal network have led to a "cascade-of-control" model examining multiple levels of Stroop interference and cognitive control \citep{banich2019stroop}. The posterior regions of the lateral prefrontal cortex regions bias processing towards task-relevant and away from task-irrelevant information, the caudal mid-cingulate regions are involved in response selection, and the anterior cingulate cortex (ACC) sends feedback to the lateral prefrontal cortex based on response accuracy.

\subsection{Computational Models}

The Stroop effect has been replicated computationally through various neural network (NN) models of automaticity and control processes. In one such approach, parallel distributed processing, word reading and color naming pathways run simultaneously with different connection strengths --- word reading has stronger connections due to habitual reading while color naming is weaker \citep{Cohenparallelprocessing}. Task demand units modulate attention in these pathways by enhancing processing in one pathway while reducing activation of the other. Normally, the stronger word reading connections result in correct naming with slower reaction times in the incongruent condition, but increased attention in the color naming units can be used to induce errors. 

This model was used to study sensory input factors such as color types. Using trichromatic processing as an input layer, opponent color pairs (blue-yellow as compared to non-opponent color pairs such as blue-red) resulted in reduced interference in incongruent naming conditions \citep{Stroopindividualdiffs}.
The Cohen model was extended with top-down conflict monitoring, where the unit’s activation increases depending on the frequency of past incongruent trials \citep{botvinick2001conflict}. Following an error, the responses were slower and more accurate. The simulations revealed interference both at the perceptual encoding stage of color processing in the Stroop task, as well as from conflict monitoring at the response level.

This paper builds on existing computational models by implementing self-organizing maps (SOMs) to study the emergence of the Stroop interference from the model structure’s competitive learning mechanisms itself, without
explicit top-down control.

\section{Approach}

The Stroop effect is observed in a color naming task, where concurrent linguistic input either competes with or coincides with the visual input. Color naming is a function of the human lexicon, the part of the brain concerned with representing and translating between word forms and their meanings. Thus, to gain insight into the Stroop effect, this paper builds on BiLex, a neural network model of the human lexicon \citep{PENALOZA2019104643,kiran2013a, grasemann-ICCM-2019, grasemann_scirep}. This section reviews BiLex, extends it with lateral connections to model response times, describes how the model was trained, and how the Stroop effect was measured on it.


\subsection{The BiLex Model}

BiLex consists of self-organizing maps (SOMs; \cite{Kohonen1982, kohonen2001}) that learn to organize word forms and their meanings according to semantic and lexical similarity. Maps are connected by learned associative connections that are used to translate between alternative semantic or lexical representations of a given word. As an example of a lexical task, picture naming is simulated in BiLex by presenting a semantic word representation to the semantic map and propagating the resulting map activation to a lexical map, which then uses its activation to generate lexical output. 

\begin{figure}[t!]
  \centering
  \includegraphics[width=\columnwidth]{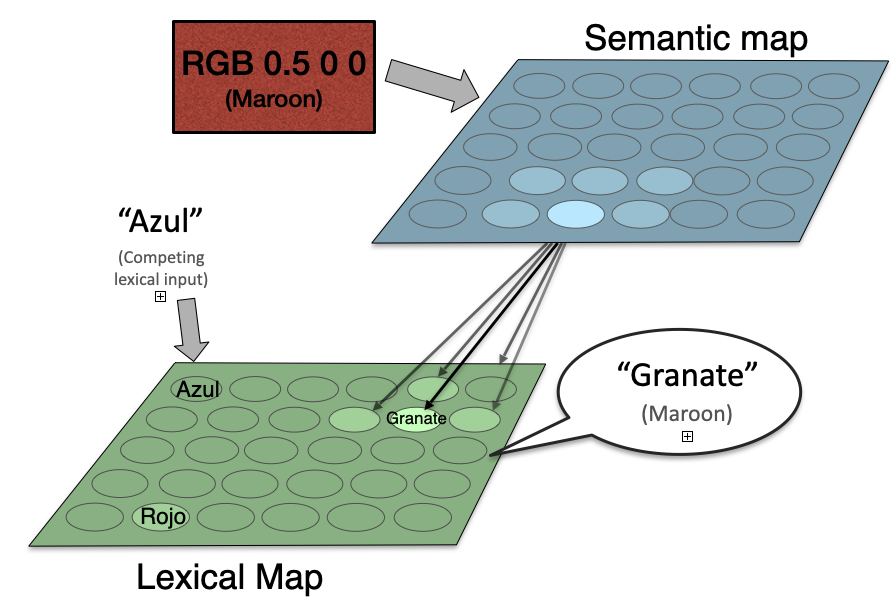}
  \vspace*{-1ex}
  \caption {The BiLex model of the Stroop effect. The semantic color input and concurrent lexical input are combined in the lexical map, and a lexical output is generated naming the color (maroon in this case). When the inputs are congruent, the response is fast and reliable; when they are incongruent, slower and error-prone. The errors can be seen as a cost of optimization of overall response times. \vspace*{-2ex}}
  \label{fg:bilex}
\end{figure}

The original BiLex was designed to model Spanish/English bilingual subjects, and therefore had one semantic map and two lexical maps, both modeling spoken words (i.e.\ phonetic maps). For the Stroop effect, only one lexical map is needed; Spanish was used for the current study. For simplicity, it was used to represent both the concurrent linguistic input and the task output. Figure \ref{fg:bilex} illustrates the architecture of the model.

\subsection{Extension with Lateral Connections}

The original BiLex was based on traditional SOMs, where map activation and learning are governed by a predefined neighborhood function, centered around the maximally responding map unit (see Table~\ref{tab:Parameters_Table} for the parameters of this model). This paper extends the SOMs with lateral connections, i.e., the map activation settles over several time steps through lateral interactions between map neurons. This settling process is an important extension of the original BiLex: it allows modeling response times, which is crucial for the Stroop effect. Inhibitory lateral connections are uniform, i.e. connection strengths between each pair of units are the same. Excitatory connections focus on a small Gaussian map neighborhood with standard deviation $\sigma$. Both the outgoing excitatory and inhibitory connection weights for each map unit sum to one.

\begin{table}[!t]
\begin{center} 
\caption{SOM Parameter definitions.}
\label{tab:Parameters_Table}
\vskip 0.12in
\begin{tabular}{ll} 
\hline
Parameter & Definition \\
\hline
    $c$   & Semantic/lexical feature vector \\
    $C$   & Input set \\
    $i$ & Map neuron \\
    $\alpha$ = 0.08 & Learning rate \\
    $\beta$ = 0.1 & Learning rate (lateral connections)\\
    $r_\mathrm{lex} = 0.45$ & Lexical input interference strength \\
    $r_\mathrm{sem} = 0.05 $ &  Semantic to lexical map influence strength \\
\hline\\[-5ex]
\end{tabular} 
\end{center} 
\end{table}

When an input $c \in C$ (a feature vector representing its semantic meaning or its lexical word form) is presented to a map, the initial activation for each map neuron $i$ is based on normalized and scaled similarity calculated  as
\begin{equation}
a_{0,i} = \mathrm{softmax}(\frac{|u_i-c|^2}{C_m T}), \label{eqn:present}
\end{equation}
where $u_i$ is the map unit $i$’s feature vector, $c$ is the map input, $C_m$ is the median Euclidean distance between pairs of vectors in the entire input set $C$, and $T$ is a temperature parameter that controls initial activation focus.

The initial activation is then updated for all units $i$ over several time steps $t$:\\[-0.5ex]
\begin{equation}
a_{t+1, i} = \alpha a_{t,i} + \beta \sum_{j}{a_{t,j}(w_{ij}^{+}-w_{ij}^{-})},\\[-0.5ex]
\label{eqn:step}
\end{equation}
where $\alpha$ controls how fast activation fades, $\beta$ controls the strength of lateral interactions, and $w_{ij}^{+-}$ are excitatory and inhibitory connections strengths between units $i$ and $j$. Over time, long-range inhibition and short-range excitation focus the map's activation on a group of neighboring similar map neurons. 

\begin{figure}[t!]
  \centering
  \includegraphics[width=\columnwidth]{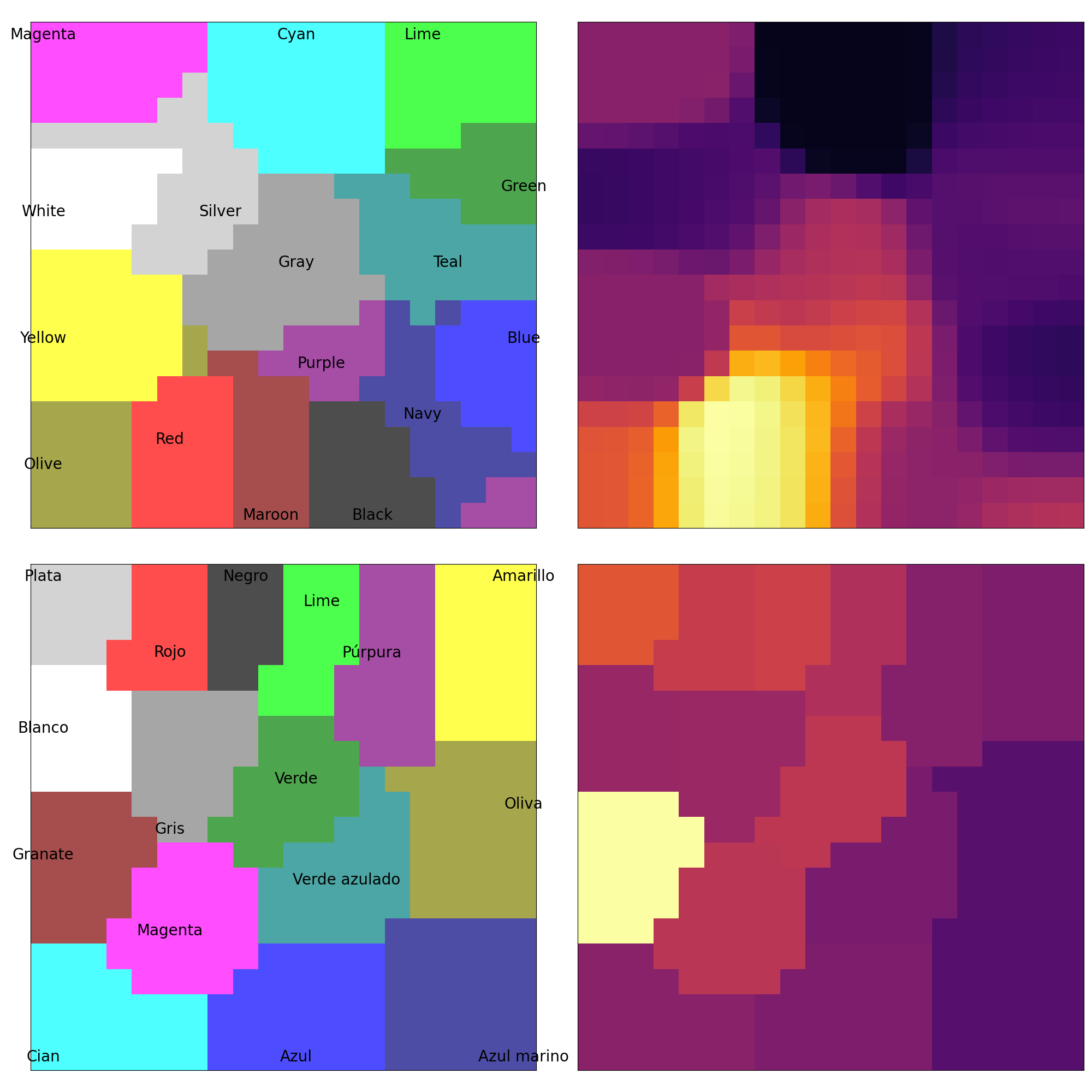}\\
  \vspace*{-1.5ex}
  \caption {The organization and activation in the BiLex model of the Stroop Effect. The model implements SOMs with lateral connections, with map units representing 16 colors and words. The semantic map (\textit{top}) was trained with RGB values of each color, and the lexical map (\textit{bottom}) with Spanish phonetic representations of words for each color. Initial map activation for the color maroon/granate is depicted in the right column. Similar colors are nearby in the semantic map, and similar words in the lexical map. As a result, the map activations are smooth and continuous.}
  \label{fg:maps}
\end{figure}

At each time $t$, the likelihood of a specific map input $c$ is calculated as\\[-2.5ex]
\begin{equation}
L(c,t) = \sum_{i \in M_c}{a_{i,t}},\label{eqn:likelihood}
\end{equation}
where $M_c$ is the set of map units $i$ that are closer to $c$ than to any other input in $C$ (shown as background colors in Figure~\ref{fg:maps}). 
Calculating $L(c,t)$ for all inputs in $C$ gives a probability distribution whose (negative) entropy $E_t$ reflects the map's uncertainty at time $t$ about which input $c$ is observed. In the reported experiments, $E_t<1$ was used as a threshold to stop lateral interactions and produce the maximum-likelihood response $c$ as the map’s output.

\subsection{Model Training}

A set of 16 words was used for model training, each representing a standard color. RGB values of each color were used as semantic representations (e.g.\ red was (1, 0, 0)), and lexical representations were generated as in previous BiLex models \citep{PENALOZA2019104643}, starting with IPA transcriptions of each color name, and transforming them into vectors of phonetic features (e.g.\ openness for vowels, place and manner for consonants).

Maps were trained similarly to standard SOMs, i.e.\ by repeatedly presenting the map with a randomly chosen word $c$, and adjusting a map neighborhood to become more similar to the input. However, instead of using a neighborhood surrounding the map unit most similar to the input, the map was presented with the input pattern (Eqn.~\ref{eqn:present}), and then updated for 30 time steps (Eqn.~\ref{eqn:step}). Each unit's feature vector $u_i$ was then adjusted to become more similar to the input $c$, with more highly activated units changing more: 
\begin{equation}
u_i' = (1-\eta a_i)u_i + \eta a_i c,
\end{equation}
where $\eta$ is the map's learning rate. Over time, as with regular SOMs, this process results in neighboring units becoming more similar to each other, and the map as a whole representing the input vectors more and more closely (as shown on the left side of Figure~\ref{fg:maps}).

Both maps in the experiments consisted of 20x20 units. They were trained for 1000 epochs, with the learning rate decreasing exponentially from 0.2 initially to 0.05 at the end of training. The standard deviation $\sigma$ of excitatory lateral connection weights decreased from 2.0 to 0.25 at the same time.    

Associative connections between the semantic and lexical map were trained with Hebbian learning. Both maps were repeatedly presented with the same randomly chosen input color, and the resulting map activations were updated until the map's entropy fell below one bit. Connections between semantic unit $i$ and lexical unit $j$ were then increased if they responded to the same input:
\begin{equation}
w_{ij}' = w_{ij} + \theta a_i a_j,    
\end{equation}
where $\theta$ is the learning rate for associative connections. In this way, co-activated neurons were connected more strongly over time. Associative connections were trained for 10 epochs; the learning rate $\theta$ was set to 0.1.  After each epoch, associative connections were normed such that each neuron’s output weights summed to one. 

\subsection{Simulating the Stroop Task}

The Stroop task can be simulated in the model as follows:
\begin{enumerate}
\item The semantic representation of a color (its RGB values) is presented to the semantic map (Eqn.~\ref{eqn:present}), resulting in an initial map activation like the one for "Maroon" shown in the top right panel of Figure~\ref{fg:maps}. 

\item A congruent or incongruent lexical input is presented to the lexical map, as shown in the bottom right panel of Figure~\ref{fg:maps}. In the 'no input' condition, no lexical input is presented to the lexical map. The strength of the interference (referred to below as \textit{$r_\mathrm{lex}$}) can be varied by multiplying the initial activation by a number between 0 and 1, where 0 reflects no interference.

\item Map activations for both maps are updated using lateral interactions between map units (Eqn.~\ref{eqn:step}), resulting in incrementally more focused activations.

\item Activation is transferred from the semantic to the lexical map through associative connections:
$a_{t,j} = a_{t,j} + \gamma \sum_i a_{t,i} w_{i,j},$ where $w_{i,j}$ is the associative connection weight between semantic unit $i$ and lexical unit $j$, and $\gamma$ determines the strength of between-map interaction.

\item If the entropy $E_t$ of the lexical map is below a given threshold, the maximum-likelihood lexical output $c$ (according to Eqn.~\ref{eqn:likelihood}) is produced as the color named by the model. If the maximum number of steps is exceeded, naming fails. Otherwise, return to Step~2. 
\end{enumerate}

The Stroop task simulation is counted as successful if the output color produced corresponds to the one presented to the semantic map, independent of the competing lexical input. The number of time steps $t$ is recorded as the time required to complete the Stroop task. A complete factorial design with 16 colors was used to evaluate the congruent (16 input cases), no input (16 input cases), and incongruent conditions (240 input cases).  

An important part of setting up the model is to determine the right level of interference, i.e.\ the relative strength of routing the semantic and congruent lexical input.
Optimal routing parameters were found by running the Stroop task simulation with a baseline  $r_\mathrm{sem}$ value of 0.05 and varying the $r_\mathrm{lex}$ value from 0.0 to 1.0 in 0.05 increments. Error rates and response times across all three conditions were recorded. The quality $q$ of each $r_\mathrm{lex}$ value was measured as
\begin{equation}
\footnotesize
q = \frac{e_i - \min(e_i)}{\max(e_i) - \min(e_i)} +
    \frac{t_i - \min(t_i)}{\max(t_i) - \min(t_i)} +
    \frac{t_c - \min(t_c)}{\max(t_c) - \min(t_c)}
\end{equation}
where $e_i$ = mean incongruent error rate, $t_i$ = mean incongruent response time, $t_c$ = mean congruent response time (there were no errors in the congruent condition and therefore $e_c$ was not needed). The min and max indicate the smallest and largest observations of these values; min-max normalization converts $e_i$, $t_i$, and $t_c$ to values between 0 and 1. Note that a low $q$ score can be achieved by a very good performance in one metric, even when the performance is poor in the others. The $q$ score thus balances the need to respond fast and to reduce errors at the same time.
The  $r_\mathrm{sem}$ and $r_\mathrm{lex}$ values with the best $q$ value were chosen for the rest of the experiments.

In the experiments, the threshold for $E_t$ was always one bit. The strength $\gamma$ of between-map interactions was 0.05. The temperature $T$ in Eqn.~\ref{eqn:present} was 0.7 for the semantic map and 0.5 for the lexical map. $\alpha$ and $\beta$ in Eqn.~\ref{eqn:step} were 0.8 and 0.1. The maximum number of time steps was set to 500, which was never reached in the reported experiments. All parameters were set empirically. Linear mixed model analysis with the semantic target color as a repeated measure was conducted to determine whether the response time differences were statistically significant between the Stroop conditions.

\section{Results}

\begin{figure}[t!]
  \centering
  \includegraphics[width=\columnwidth]{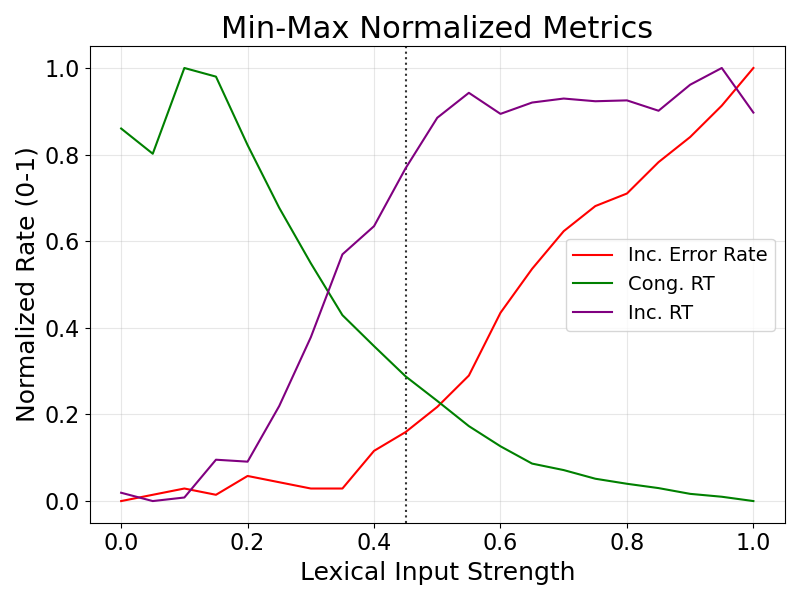}\\
  \vspace*{-3ex}
  \caption {Finding the best values for the $r_\mathrm{lex}$ routing parameter. The normalized $e_i$ (\textit{red}), $t_i$ (\textit{purple}), and $t_c$ (\textit{green}) components of the $q$ metric are depicted for each possible $r_\mathrm{lex}$ value. The optimal routing parameter was found to be an $r_\mathrm{lex}$ of 0.45, marked by the dashed black line, balancing color naming speed with minimized naming errors.}
  \label{fg:routingparams}
\end{figure}

Lower $r_\mathrm{lex}$ values resulted in slower response times and reduced incongruent errors, while higher $r_\mathrm{lex}$ values allowed completing the color naming task faster, but with more incongruent errors. Normalized values for $e_i$, $t_i$, and $t_c$ were compared. Lower $r_\mathrm{lex}$ values, like between 0.30 and 0.40, have lower $e_i$ normalized values, but higher $t_i$. Higher $r_\mathrm{lex}$ values, e.g.\ 0.50 to 0.65, have higher $e_i$ and $t_i$ normalized values, but low $t_c$. Further, the max-min spread for 0.45 is smaller compared to other $r_\mathrm{lex}$ values. The best routing parameters were found to be $r_\mathrm{lex}$=0.45 and  $r_\mathrm{sem}$=0.05, as max-min spread is smaller compared to other $r_\mathrm{lex}$ values, indicating balanced performance (Figure~\ref{fg:routingparams}).

The simulations demonstrated a strong Stroop effect. Overall, across all three conditions, the model named the colors correctly 84.2\% of the time. When presented with congruent inputs, the model was 100\% accurate, with incongruent inputs, 83.0\% accurate, and with no-input cases, 87.5\% accurate.

\begin{figure}[t!]
  \centering
  \includegraphics[width=\columnwidth]{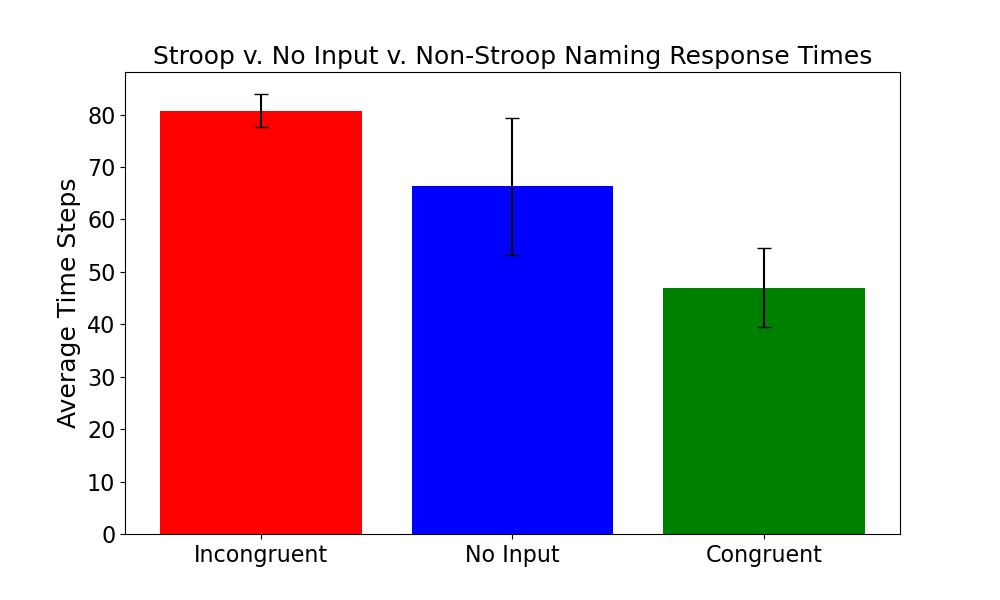}\\
  \vspace*{-3ex}
  \caption {The mean and standard error (SE) of RTs across the three conditions of the Stroop effect: congruent (\textit{red}), no-input (\textit{blue}), and incongruent (\textit{green}). RTs improve significantly for congruent vs.\ no-input vs.\ incongruent conditions, indicating that the model exhibits a strong Stroop effect.}
  \label{fg:responsetime}
\end{figure}

Similar results were observed in response times as well. The linear mixed model analysis used 229 successful congruent, incongruent, and no-input cases across the 16 target color groups. Group sizes ranged from 2 to 17 cases, with a mean group size of 14.3. Compared to congruent input, the incongruent and no input conditions significantly delayed the output convergence (Figure~\ref{fg:responsetime}; $p < 0.001$). The incongruent condition had the strongest effect, with accurate color naming taking approximately 55 steps longer than the congruent condition ($\beta = 55.05, \mathrm{SE} = 6.08, 95\%~\mathrm{CI}=[43.1, 67.0]$). The no-input cases also converged more slowly, on average taking 36 steps longer than the congruent condition ($\beta = 35.87, \mathrm{SE} = 8.58, 95\%~\mathrm{CI}=[19.1, 52.7]$). Group variance (3485.2) indicated that there was high variation between different semantic inputs. However, the Stroop effect within the combined model persisted across all input patterns for the same task, with statistically significantly slower response times for incongruent vs. no input vs. congruent conditions.

\begin{figure} [t!]
            \centering
            \includegraphics[width=\columnwidth]{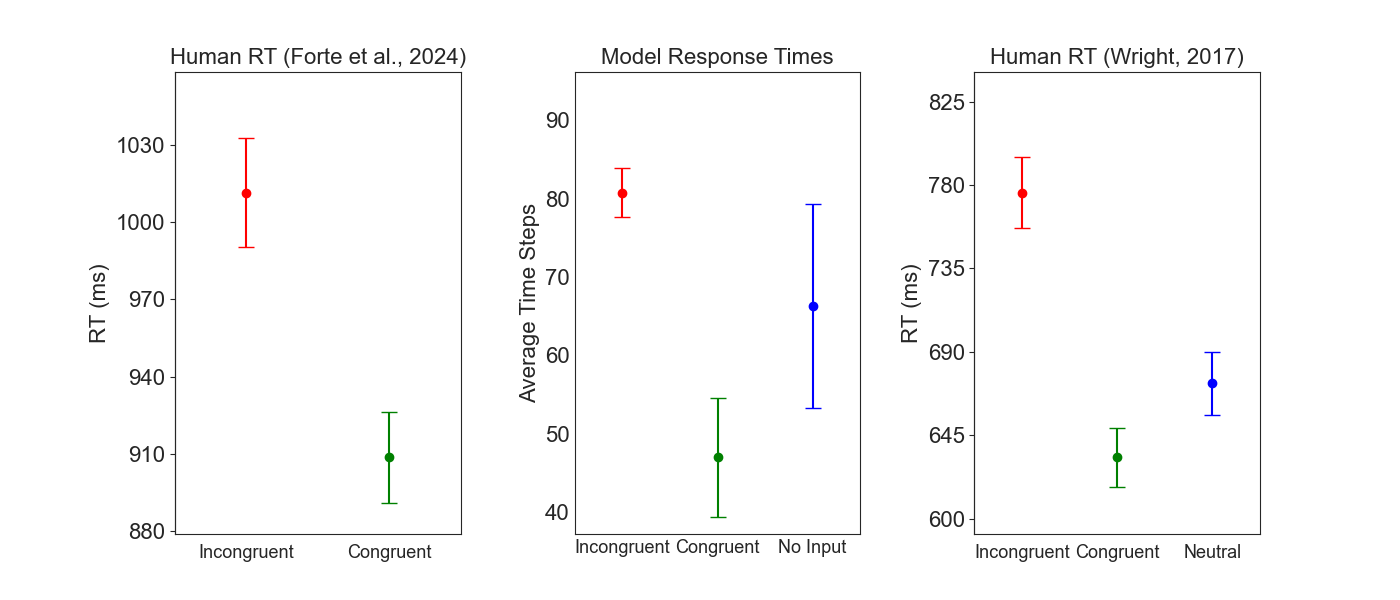}
            \vspace*{-3ex}
            \caption{The mean and SE of RTs in the model compared with human performance. Left: n = 98, ages: 37 - 55 years \citep{forte2024stroopdistribution}; right: n = 77, ages: 17 - 45 years \citep{Wright2017}. The vertical scales were adjusted to line up the averages; the SEs are mostly similar and the differences between the conditions are significant, suggesting that the model replicates the human data.}
            \label{fig:rt_distribution_comparisons}
        \end{figure}

Figure~\ref{fig:rt_distribution_comparisons} compares model response times to those of human adults in two prior studies \citep{forte2024stroopdistribution, Wright2017}, where mean reaction times (RTs, in ms) were measured with a computerized Stroop test. Results indicate that the model performs similarly to humans. The absolute RT values and variance are different, but in all cases there is a statistically significant differences between conditions.

\begin{figure}[t!]
  \begin{center}
  \includegraphics[width=\columnwidth]{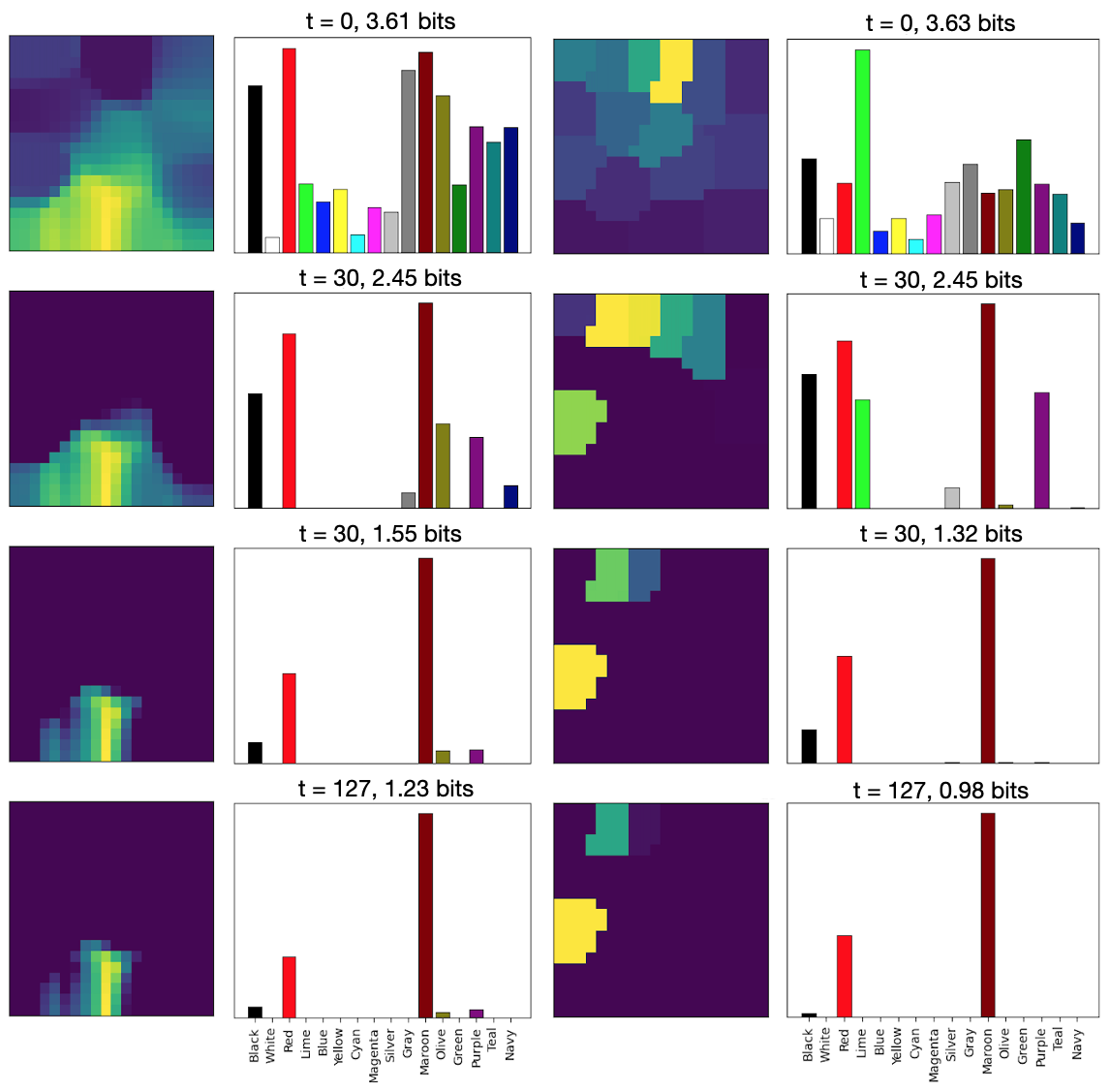}\\
  \end{center}
  \vspace*{-3ex}
  \caption {The response of the model over time. Lexical map activation (\textit{right}) is shown over time for incongruent inputs. Initially (at t = 0), map activation (\textit{left}) only reflects the lexical input. However, as time progresses, the correct input (maroon) is transferred from the semantic to the lexical map and gains more weight (t = 30), eventually causing the lexical map to converge on the correct output (t = 127). It is this competition over time, mediated by the lateral connections, that results in the Stroop effect.}
  \label{fg:incongruenttime}
\end{figure}

\begin{figure}[!t]
  \centering
  \includegraphics[width=\columnwidth]{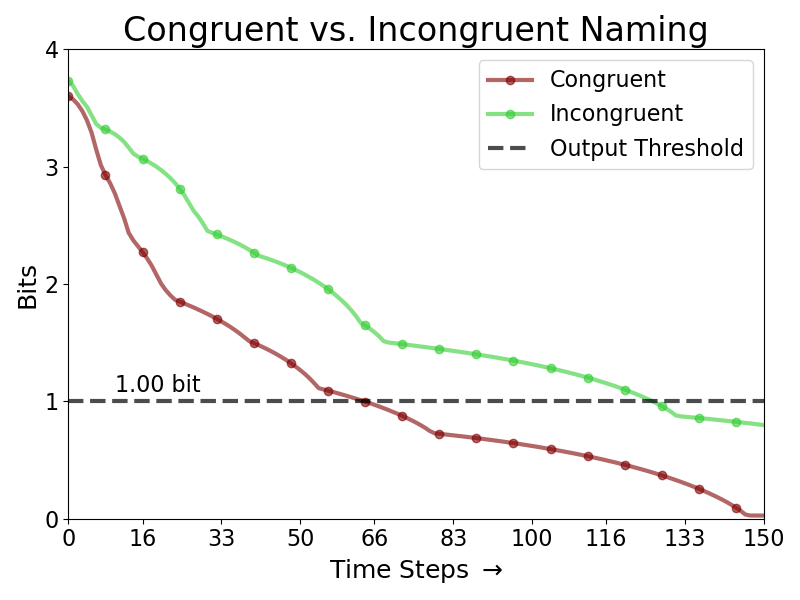}
  \vspace*{-6ex}
  \caption {Comparing the model response to the congruent and incongruent inputs. When the lexical map is given incongruent input (\textit{lime}), the map takes significantly longer to focus on the specified color and cross the output threshold for uncertainty below 1.0 bit than when the inputs are congruent (\textit{maroon}).}
  \label{fg:incongvscong}
\end{figure}

Figure~\ref{fg:incongruenttime} illustrates the progression of the lexical and semantic map activations in an incongruent input case. Over time, activation is transferred from the semantic map to the lexical map, enabling the lexical map to converge on the correct color output in this case. Figure~\ref{fg:incongvscong} illustrates the responses over time for incongruent and congruent naming, demonstrating that the rate of convergence on the correct output is faster in the congruent condition.

\section{Discussion and Future Work}

The main result is that the laterally connected BiLex model can account for the Stroop effect. Incongruent inputs take longer to process and result in more errors than congruent inputs. The same is true to a lesser extent for the no-input condition. Importantly, this process arises from the interactions within and between the self-organizing maps, mediated by the lateral and associative connections. Attention mechanisms are implicitly represented through these interactions, demonstrating how performance is optimized.

The current experiments should be improved by modeling transfer in semantic conflict and response conflict effects \citep{CHEN2013577}, stimulus onset asynchrony (SOA), word reading vs.\ color naming performance, and practice effects. It may also be possible to model age-related sensitivity, i.e.\ how the Stroop effect becomes stronger as a result of a reduced capacity for task-inhibition in older age \citep{MILHAM2002277}, or simulate performance in patients with impairments by systematically lesioning the semantic and lexical maps, as implemented with BiLex models of aphasic patients \citep{BiLexAphasia}.

The routing parameters that were found to maximize naming speed while minimizing error responses reveal an interesting effect. With competing lexical input, both naming errors and response time increase. , while blocking competing input leads to more accurate and faster responses. Naming is faster on average when the lexical input and transfer strength are above zero; however, the cost of this improvement is that the Stroop effect emerges with incongruent cases. Data-driven selective attention towards the most salient inputs is a crucial part of human cognition \citep{algom2019reclaiming}; the Stroop effect thus arises from the intention to use available inputs optimally.  By optimizing lexical input strength (\textit{$r_\mathrm{lex}$} = 0.45), the model can achieve full accuracy in the congruent condition and minimize error in the incongruent condition, while achieving an efficient run time. Given that inputs in the real world are more likely to be congruent, Stroop errors can be seen as a natural side effect of a system that performs well in most cases.

In the current model, routing was simply reduced to two parameters, but it is a complex mechanism in its own right. A more accurate modeling of routing in BiLex could lead to further insights into cognitive control. One promising approach is the conditional routing model \citep[CRM;][]{stocco2010conditional}, in which the basal ganglia implement an adaptive information routing system that modulates signals between pairs of cortical regions. Similar to CRM, the routing rules in BiLEx can be learned through contrastive Hebbian learning \citep[CHL;][] {dayanabbot2001, Rolls1998NeuralNA}, which uses only locally available information to adapt connections and represent routing rules within the striatal part of the network.

While the original CRM executes one interaction between a pair of cortical regions at a time, the new model would continually update (and possibly reevaluate) all routing parameters within the lexicon, creating an adaptive, task-dependent dynamic system rather than a binary routing mechanism. Unlike previous
bilingual Stroop models \citep{yoon2011computational}, the new model would demonstrate how bilingual lexical function in humans is optimized over time. It is therefore ideally suited to explain how bilingual Stroop interference arises from stimulus-driven attention combined with top-down control. In addition to the Stroop effect, BiLex with conditional routing can help understand language acquisition and lexical access in L1 and L2, priming effects, response timing, true and false cognates, and language and task switching effects. This expansion follows logically from the Stroop effect experiments in this paper and constitutes a most compelling direction of future work.

\section{Conclusion}

The BiLex self-organizing map model of the lexical/semantic system was adapted to simulate the Stroop effect. When extended with lateral connections, the model could account not only for the increased errors in the incongruent condition but also for the differences in response time between congruent, no-input, and incongruent conditions. Interestingly, the Stroop effect in the model can be seen as a necessary cost for optimizing overall response times with accuracy. The model can thus serve as a foundation for studying cognitive control and the effects that arise from it.
    
\section{Acknowledgments}
The authors wish to thank Andrea Stocco and Arturo Hernandez for valuable discussions. This research was supported in part by NIH under grant 5R01DC020653-02.

\bibliographystyle{apacite}

\setlength{\bibleftmargin}{.125in}
\setlength{\bibindent}{-\bibleftmargin}

\bibliography{CogSci_Template}

\end{document}